\documentclass{PoS}

\title{Spontaneous chiral symmetry breaking on the lattice}

\ShortTitle{Spontaneous chiral symmetry breaking on the lattice}

\author{\speaker{Shoji Hashimoto}\\
        High Energy Accelerator Research Organization (KEK), 
        Tsukuba 305-0801, Japan.\\
        E-mail: \email{shoji.hashimoto@kek.jp}}

\abstract{
  Using lattice QCD we study the spectrum of low-lying fermion
  eigenmodes.
  According to the Banks-Casher relation, accumulation of the low-mode
  is responsible for the spontaneous breaking of chiral symmetry in
  the QCD vacuum.
  On the lattice we use the overlap fermion formulation that
  preserves exact chiral symmetry.
  This is essential for the study of low-lying eigenmode distributions.
  Through a detailed comparison with the expectations from chiral
  perturbation theory beyond the leading order, we confirm the senario
  of the spontaneous symmetry breaking and determine some of the low
  energy constants.
  We also discuss on other related physical quantities, which can be
  studied on the lattice with exact chiral symmetry.
}

\FullConference{6th International Workshop on Chiral Dynamics\\
		 July 6-10 2009\\
		 Bern, Switzerland}

\begin{document}

\section{Introduction}

Spontaneous breaking of chiral symmetry is the most fundamental
property of the vacuum of Quantum Chromodynamics (QCD). 
Once we assume that the chiral symmetry is spontaneously broken, we
can derive many important relations in the phenomenology of strong
interaction, such as the GMOR relation, Goldberger-Treiman relation,
and other soft pion theorems, which are written in the language of
chiral effective theory.
The problem of how and why the spontaneous chiral symmetry breaking
occurs remains a difficult question due to the non-perturbative dynamics of
QCD.

Chiral symmetry of course plays a key role in the understanding
of chiral symmetry breaking.
In the flavor non-singlet sector of chiral symmetry, pion arises as
the Nambu-Goldstone boson associated with the spontaneous symmetry
breaking, while in the flavor-singlet sector the chiral symmetry is
violated by the axial anomaly and is related to the topology of
non-Abelian gauge theory.
There are near-zero modes of quarks associated with the topological
excitations; their accumulation in the vacuum leads to the
symmetry breaking in the flavor non-singlet sector as indicated by the
Banks-Casher relation. 
Therefore, the initial setup to study the chiral symmetry breaking
should preserve both the flavor singlet and non-singlet chiral symmetries.

Lattice QCD is the most promising approach to solve the low-energy
dynamics of QCD, but there is a problem in realizing the chiral
symmetry on the lattice.
The conventional Wilson-type fermions violate the chiral symmetry at
the action level, and 
the discrimination between the physical effect of symmetry breaking
and the lattice artifact cannot be done in a clear manner.
On the other hand, the staggered fermions have a chiral symmetry but
break the flavor symmetry.
With these lattice fermions, the continuum limit has to be taken
before analyzing the data with the continuum chiral effective theory. 

Among other physical quantities, we are interested in extracting the
chiral condensate $\langle\bar{q}q\rangle$, which is an order
parameter of the chiral symmetry breaking.
This is not easy because the scalar density operator $\bar{q}q$ has a
power divergence of the form $m_q/a^2$ as the cutoff $1/a$ goes to
infinity, hence the massless limit has to be taken to obtain physical
result.
(When the chiral symmetry is violated from the outset as in the
Wilson-type fermions, the divergence is even stronger $1/a^3$.)
The condensate however vanishes in the massless limit,
as far as the space-time volume is kept finite.
Therefore, the proper order of the limits is to take the infinite
volume limit first and then the massless limit, which is
called the thermodynamical limit.
Thus, the study of symmetry breaking is numerically so demanding, and
some theoretical guidance is required to control the limits.
In QCD, the chiral perturbation theory (ChPT) provides such a
theoretical framework. 

This work is an attempt to simulate the QCD vacuum on the lattice with
exact chiral symmetry.
We use the overlap fermion formulation
\cite{Neuberger:1997fp,Neuberger:1998wv}, that exactly preserves
chiral and flavor symmetries at finite lattice spacing 
and correctly reproduces the axial anomaly. 
We calculate the chiral condensate in various ways, which provides a
good test of the chiral effective theory.
In particular, we study the spectral density of the Dirac operator and
compare it with ChPT at the next-to-leading order (NLO) of the chiral
expansion.
We also discuss a few other consequences of spontaneous symmetry
breaking and related lattice calculations, which are made possible
with exact chiral symmetry on the lattice.

These works have been done by the JLQCD and TWQCD collaborations. 
An overview of their recent physics results is found in
\cite{Hashimoto:2008fc}.

\section{Dirac spectrum and chiral symmetry breaking}
The chiral symmetry breaking is induced by an accumulation of
low-lying eigenstates of quark-antiquark pair, as indicated by the
Banks-Casher relation \cite{Banks:1979yr}
\begin{equation}
  \label{eq:Banks-Casher}
  \lim_{m\to 0}\lim_{V\to\infty}\rho(\lambda=0) = \frac{\Sigma}{\pi},
\end{equation}
where $\rho(\lambda)$ denotes the eigenvalue density of the Dirac
operator, 
$\rho(\lambda)\equiv
(1/V)\sum_k\langle\delta(\lambda-\lambda_k)\rangle$.
The expectation value $\langle\cdots\rangle$ represents an ensemble
average and $k$ labels the eigenvalues of the Dirac operator on a
given gauge field background.
On the right hand side of (\ref{eq:Banks-Casher}), 
$\Sigma$ is the chiral condensate, $\Sigma=-\langle\bar{q}q\rangle$,
evaluated in the massless quark limit.
In the free theory, we expect a scaling $\rho(\lambda)\sim \lambda^3$,
which vanishes at $\lambda=0$, for a dimensional reason.
The relation (\ref{eq:Banks-Casher}) implies that the spontaneous
chiral symmetry breaking characterized by non-zero $\Sigma$ is
related to the number of near-zero modes in a given volume after taking
the thermodynamical limit. 

Based on ChPT, more detailed forms of $\rho(\lambda)$ at finite
$\lambda$, $V$ and $m$ have been obtained.
This is achieved by evaluating the chiral condensate at imaginary
value $i\lambda$ of valence quark mass, relying on the analytic
continuation.
The spectral function is thus obtained without explicitly treating the
quark eigenstates.

\begin{figure}[tb]
  \centering
  \includegraphics*[width=7cm]{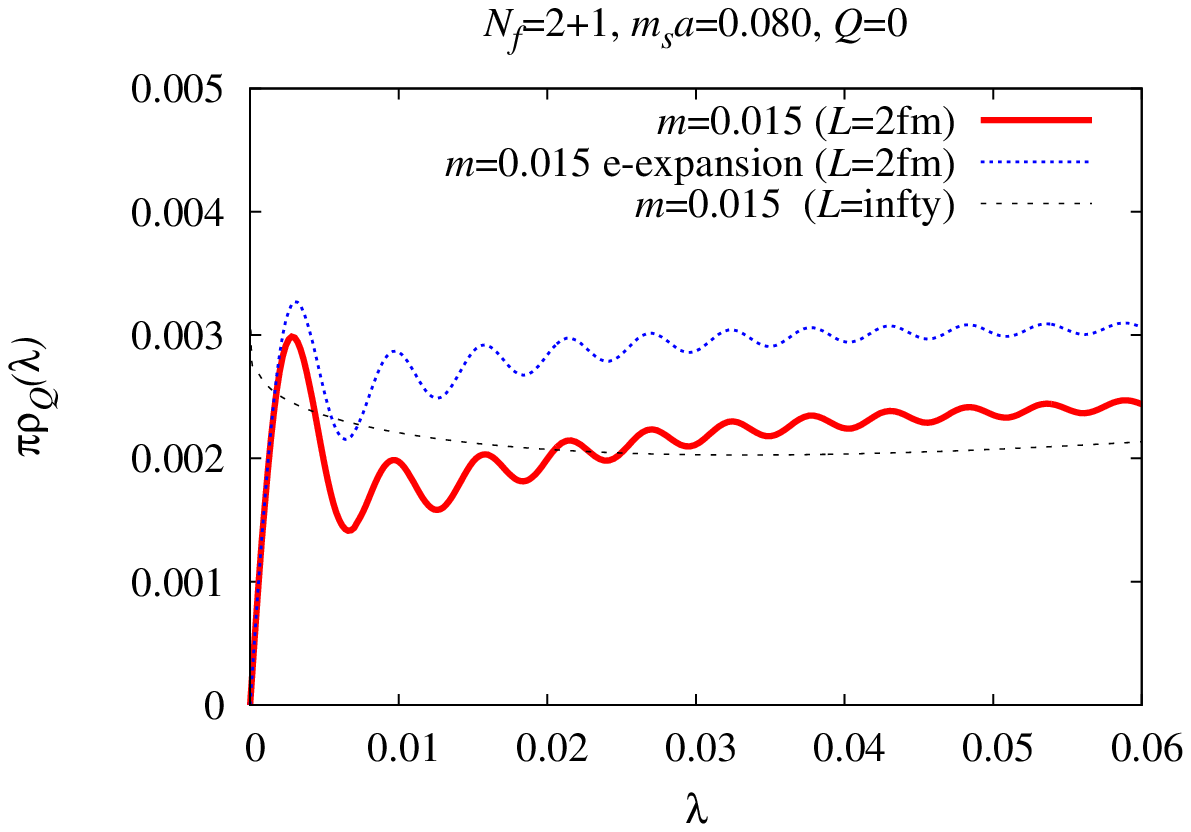}
  \includegraphics*[width=7cm]{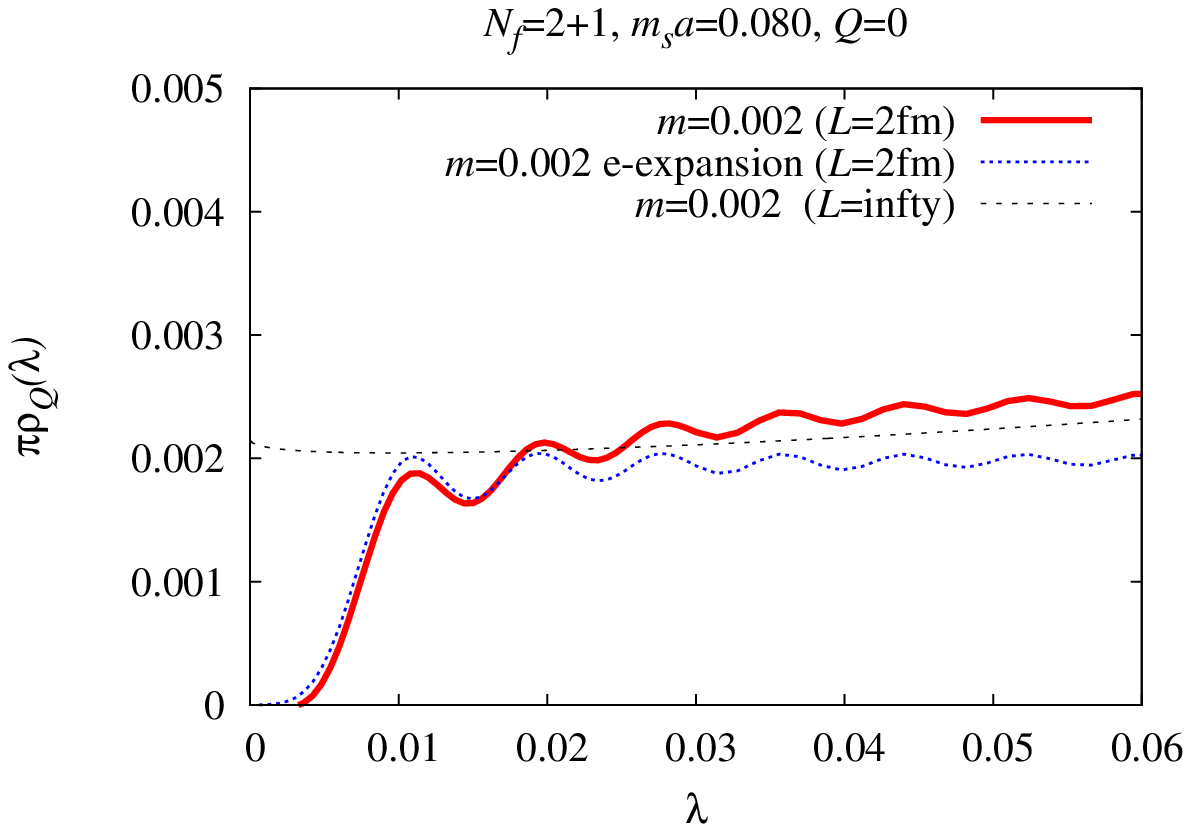}
  \caption{
    Theoretical expectation from \cite{Damgaard:2008zs} on the
    spectral function at $L\sim$ 2~fm.
    The left and right panels correspond to the pion mass around
    300~MeV and 100~MeV, respectively.
    The curves correspond to the full NLO calculation (solid curve,
    red), the leading order result in the $\epsilon$-expansion (dashed
    curve, blue), and the result in the infinite volume (dotted curve,
    black) at NLO of the conventional $p$-expansion.
  }
  \label{fig:Damgaard-Fukaya}
\end{figure}

In this work, we use the most recent calculation by Damgaard and
Fukaya \cite{Damgaard:2008zs}, which is based on the conventional
$p$-expansion of ChPT but includes an integral over zero-momentum modes
of pion field, hence it is also consistent with the $\epsilon$-expansion.
At NLO, they provide a formula for $\rho(\lambda)$ at finite $V$ and
(non-degenerate) $m$.
A typical example is shown in Figure~\ref{fig:Damgaard-Fukaya} (left panel).
The plot shows the spectral function $\rho(\lambda)$ at a given volume
$V=L^3\times(3L)$ with $L\sim$ 2~fm and pion mass $m_\pi\sim$ 300~MeV.
Within the leading order (LO) of the $\epsilon$-expansion, the spectral
function is given by the dashed curve, which is equivalent to what one
obtains from the chiral random matrix theory.
It receives a finite volume correction at NLO and becomes the solid
curve.
Starting from the second peak of the oscillating curve, the
spectral function is suppressed due to the pion-loop effects.
In the infinite volume limit, we expect a smoother dotted curve, which 
contains a chiral logarithm.

The right panel of Figure~\ref{fig:Damgaard-Fukaya} shows the spectral
function in the $\epsilon$-regime.
Here the lowest-lying eigenvalue is strongly suppressed by the fermion
determinant as the quark mass is close to zero.
The difference between LO and NLO in the $\epsilon$-expansion is less
significant, since the system is in the $\epsilon$-regime.

Once we could calculate the spectral density on a finite volume
lattice, we can extract $\Sigma$ and $1/F^2$.
Essentially, $\Sigma$ determines the height of the distribution
and $1/F^2$ the size of the NLO effects.
The finite volume scaling should also be tested with more than one
lattice volumes.

\section{Lattice analysis of the spectral density}
We use the lattice data obtained in the course of dynamical overlap
fermion simulations by the JLQCD and TWQCD collaborations.
The project mainly aimed at controlling the chiral extrapolation of
physical quantities by realizing exact chiral and flavor symmetries in
the lattice simulations.
The use of the continuum ChPT is justified even at finite lattice
spacings in contrast to other lattice fermion formulations, for which
some modifications of ChPT with additional parameters is required.
With the exact chiral symmetry, there have been several new physics
applications proposed from the project \cite{Hashimoto:2008fc}.

We use the overlap-Dirac operator 
$D(0)\equiv(\rho/a)[1+X/\sqrt{X^\dagger X}]$
with the Wilson kernel $X\equiv aD_W-\rho$
\cite{Neuberger:1997fp,Neuberger:1998wv}.
This fermion formulation exactly preserves a modified version of the
chiral symmetry at finite lattice spacings \cite{Luscher:1998pq}.
The axial Ward-Takahashi identities are essentially the same as those
in the continuum theory, and the index theorem is satisfied.

The large-scale Monte Carlo simulations had been made feasible by
restricting the Markov chain in a given topological charge, since the
overlap-Dirac operator has a discontinuity on the boarder of the
topological charge, 
and the numerical costs grows as $V^2$ to treat the change of
topology.
This is achieved by introducing unphysical heavy Wilson fermions in
the simulation \cite{Fukaya:2006vs}.
Fixing topology induces a finite volume effect for any physical
quantities, but corrections are possible using the general formulae
developed in \cite{Brower:2003yx,Aoki:2007ka}. 
For the calculation of the spectral function considered in this work,
the fixed topology is an advantage rather than a disadvantage, as the
ChPT formulae are given for fixed topological sectors. 

At an early stage of the project, we performed simulations of
two-flavor QCD at a lattice spacing $a\simeq$ 0.12~fm on a 
$16^3\times 32$ lattice.
The simulation details are described in \cite{Aoki:2008tq}.
We took six values of sea quark mass in the range $m_s/6\sim m_s$,
with $m_s$ the physical strange quark mass, to investigate the chiral
extrapolation as discussed below.
We also carried out a run in the $\epsilon$-regime with the quark mass
around 3~MeV, in order to study the low-lying Dirac eigenvalue spectra,
as discussed in the next section.

We then extended the work to 2+1-flavor simulations at a lattice
spacing $a\simeq$ 0.11~fm on $16^3\times 48$ and $24^3\times 48$
lattices.
We cover a similar range of (degenerate) up and down quark masses as
in the two-flavor runs while taking two values of strange quark mass
near its physical value.
A dedicated run in the $\epsilon$-regime has also been performed with
the up and down quark masses around 3~MeV while keeping the strange
quark mass near the physical value.
A preliminary report of these runs is found in \cite{Matsufuru_lat08}.

At an early study of the eigenvalue spectrum, we use the distribution
of the lowest-lying Dirac eigenvalue to extract $\Sigma$ in two-flavor
QCD, by matching the distribution to the expectation from the chiral
random matrix theory \cite{Fukaya:2007fb,Fukaya:2007yv}.
In this method the relation to ChPT is established only at LO of the
$\epsilon$-expansion, so that the result may contain significant
finite volume effect, which is the NLO effect.
This was signalled already by a slight inconsistency between $\Sigma$
obtained from the lowest and from the second lowest eigenvalues.

With the new formula \cite{Damgaard:2008zs}, we can now consistently
incorporate the NLO effects in the analysis.
We use the 2+1-flavor data of spectral function in a wider region of
the eigenvalue $\lambda$.
Since the formula is valid also in the $p$-regime, we are able to
include the $p$-regime lattices into the analysis.

\begin{figure}[tb]
  \centering
  \includegraphics*[width=7cm]{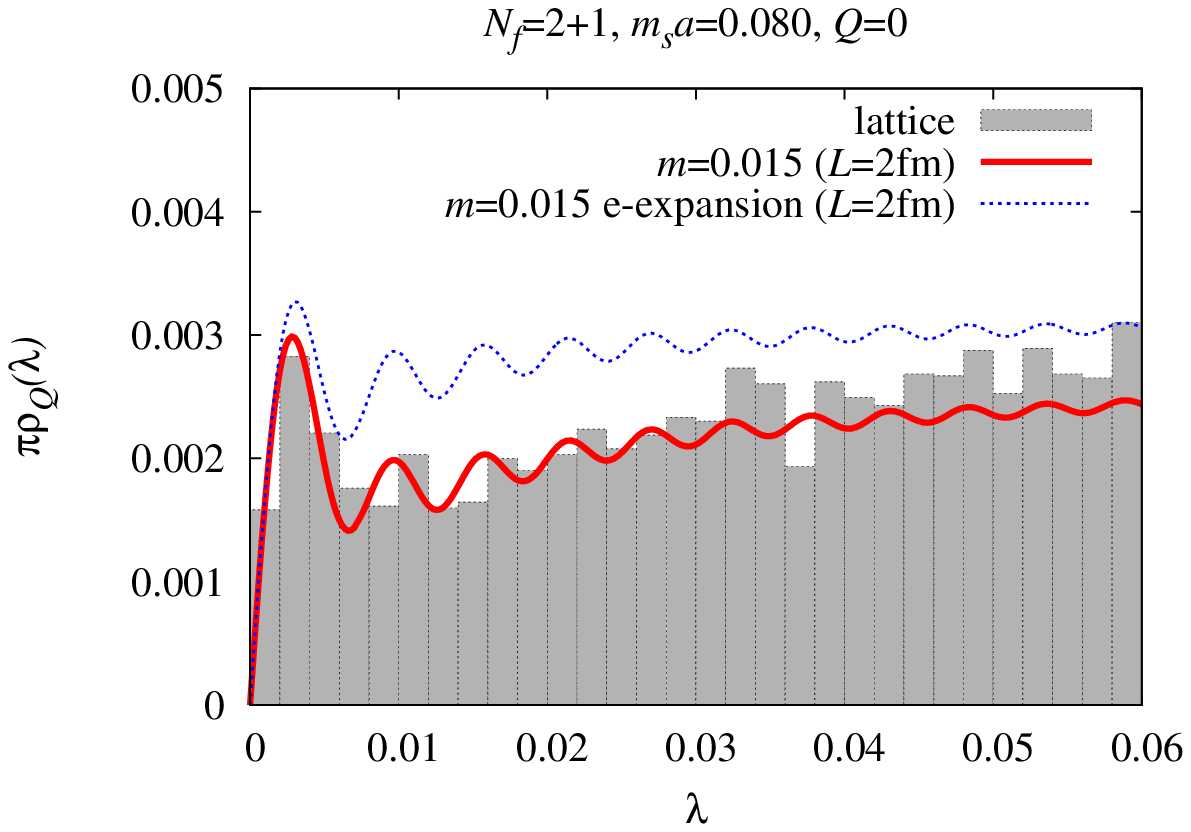}
  \includegraphics*[width=7cm]{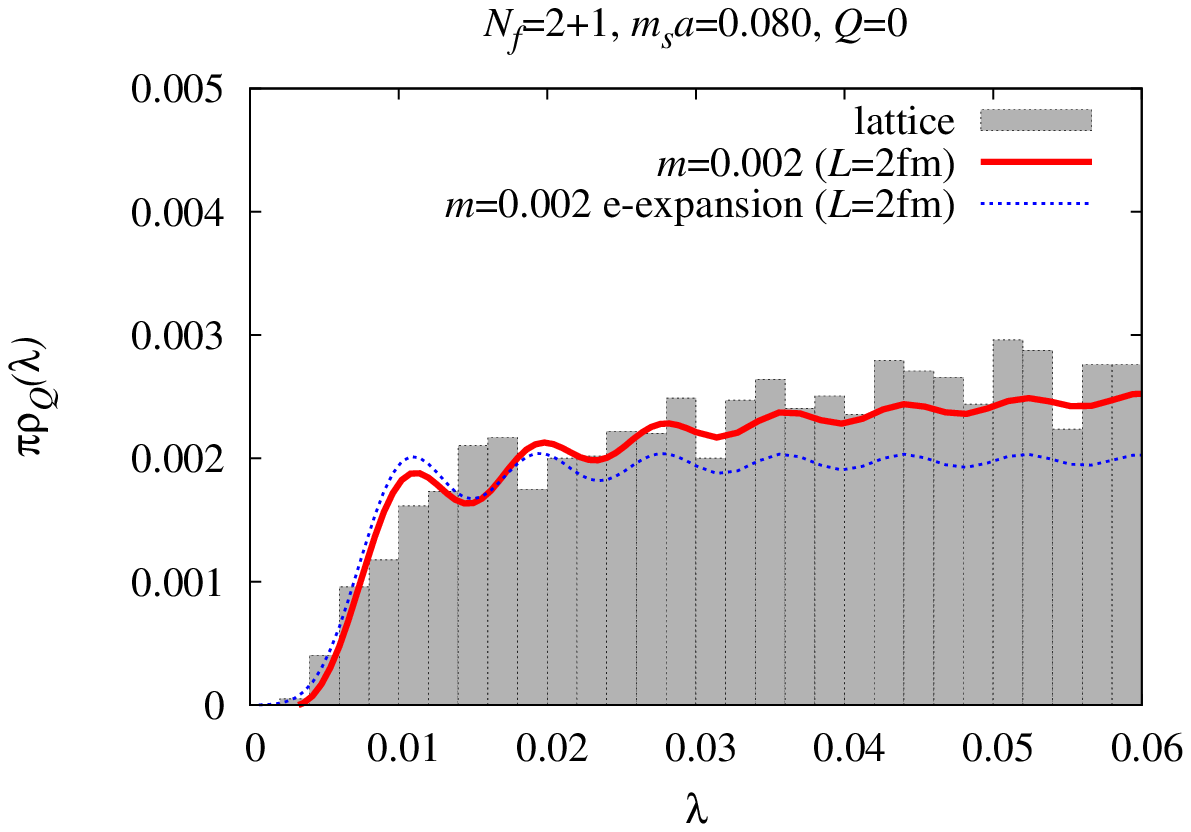}
  \caption{
    Lattice results for the spectral function (histogram) together with
    the LO (dashed, blue) and NLO (solid, red) curves.
    The left and right panels correspond to the pion mass around
    300~MeV ($p$-regime) and 100~MeV ($\epsilon$-regime), respectively.
  }
  \label{fig:spectal_func}
\end{figure}

Figure~\ref{fig:spectal_func} shows the results for the spectral
function in 2+1-flavor QCD obtained on $16^3\times 48$ lattices.
The global topological charge $Q$ is fixed to 0.
The lattice data (histogram) are overlaid on the expectation from the
NLO formula for both the $p$-regime (left panel) and $\epsilon$-regime
(right panel).
We observe that the NLO formula nicely reproduces the shape of the
lattice data.
The curves are drawn by fixing the parameters ($\Sigma$ and $F$)
with the data for an integrated spectrum at two representative values
of $\lambda$.
Typically, one is taken near the top of the first peak, and the other
is taken at $\lambda\sim$ 0.04, where the NLO effect is significant.

We have further tested the NLO formula by extending the calculation to
a larger lattice.
The finite volume scaling is tested with the data on a $24^3\times 48$
lattice in the $p$-regime.
We find that the result for $\Sigma$ from this lattice is consistent
with that on a smaller lattice after correcting the finite volume
effect for both lattices.
This implies that the lattice result scales as expected towards the
thermodynamical limit.

\begin{figure}[tb]
  \centering
  \includegraphics*[width=8cm]{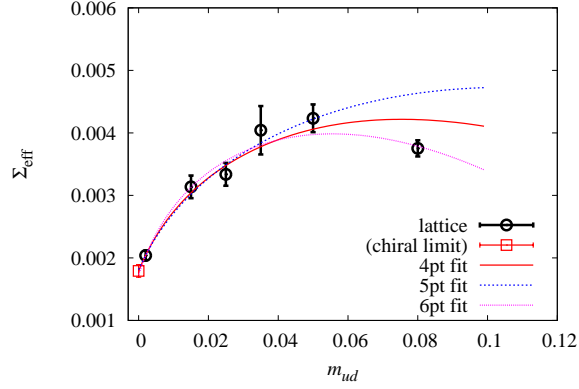}
  \caption{
    Chiral extrapolation of $\Sigma(m_{ud},m_s)$ to the limit of
    vanishing up and down quark masses.
  }
  \label{fig:sigma_chiral}
\end{figure}

From this analysis, we obtain $\Sigma$ at six values of up and down
quark masses $m_{ud}$ while keeping the strange quark mass $m_s$ at
its physical value.
Thus, we obtain an ``effective'' value of $\Sigma$ at each sea quark
mass, $\Sigma(m_{ud},m_s)$.
We extrapolate the results to the chiral limit of $m_{ud}$ using the
NLO chiral expansion in two-flavor QCD \cite{Gasser:1983yg}
\begin{equation}
  \label{eq:sigma}
  \Sigma(m_{ud},m_s)=\Sigma(0,m_s)
  \left[
    1 - \frac{3M_\pi^2}{32\pi^2 F^2}\ln\frac{M_\pi^2}{F^2}
    + \frac{32 L_6 M_\pi^2}{F^2}
  \right]
\end{equation}
as shown in Figure~\ref{fig:sigma_chiral}.
(In the actual analysis, we use the formula including finite volume
corrections.) 

Since we have the data in the $\epsilon$-regime, which is very close
to the chiral limit, the chiral extrapolation is stable against the
change of the fitting region.
We also attempt to use the 2+1-flavor formula at NLO, and find the
result in the chiral limit is consistent.

After renormalizing to the $\overline{\mbox{MS}}$ scheme at 2~GeV
using a non-perturbatively calculated renormalization factor
\cite{Noaki:2009xi}, we obtain a preliminary result
$\Sigma^{\overline{\mathrm{MS}}}(0,m_s;\mathrm{2~GeV})
= [243(4)(^{+16}_{-\ 0}) \mathrm{~MeV}]^3$,
where the second error represents an uncertainty due to the lattice
scale. 
We also obtained $F$ and $L_6$ that are consistent with phenomenological
analysis. 
For more details, see \cite{Fukaya:2009fh}.

\section{Other consequences of spontaneous symmetry breaking}
The accumulation of the near-zero modes as seen in the spectral
function is a direct measure of the spontaneous chiral symmetry
breaking, as its infinite volume limit corresponds to the Banks-Casher
relation. 
On the other hand, there are many other physical quantities that
reflect the spontaneous breaking of chiral symmetry.
In the following, we describe a few of them calculated on the same set
of lattice simulations.

\subsection{Topological susceptibility}
Topological susceptibility $\chi_t\equiv\langle Q^2\rangle/V$
characterizes how much topological excitations are occurring in the QCD
vacuum.
Although the definition involves the global topological charge $Q$,
the susceptibility itself is a local quantity and could be determined
even when $Q$ is kept fixed \cite{Aoki:2007ka}.
Namely, $\chi_t$ can be extracted from topological charge density
correlator as
\begin{equation}
  \lim_{|x|\to\infty}\langle mP(x) mP(0)\rangle_Q =
  -\frac{1}{V}\left( \chi_t - \frac{Q^2}{V} + \cdots \right)
  + O(e^{-m_{\eta'}|x|}),
\end{equation}
where $mP(x)$ is a flavor-singlet pseudoscalar density operator
related to the topological charge density through the axial-anomaly
relation. 
When $Q=0$, this correlator approaches a negative constant at large
separation $|x|$ after the excitation of the $\eta'$ meson saturates.
This is intuitively understood as follows. 
Since the global topological charge is fixed to zero, if we find a
positive topological charge excitation at a space-time point $0$, then
we have more chance to find a negative excitation at other points $x$
to keep the total to be zero.

This constant correlation is indeed observed in our work where we
calculate the flavor-singlet pseudo-scalar correlator by maximally
using the low-lying eigenmodes exactly calculated and stored on disks
\cite{Aoki:2007pw,Chiu:2008kt}.
We thus extract $\chi_t$ at each sea quark mass.

\begin{figure}[tb]
  \centering
  \includegraphics*[width=8cm]{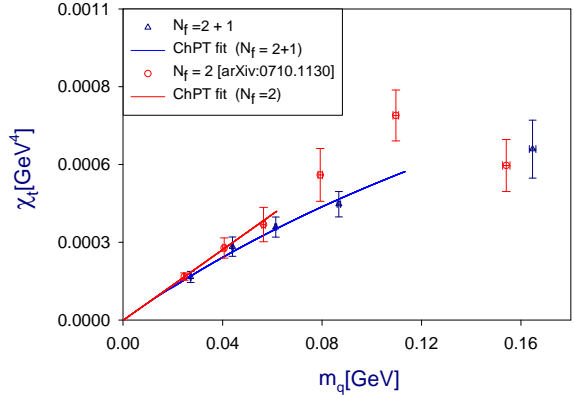}
  \caption{
    Topological susceptibility $\chi_t$ in 2 (circles) and 2+1-flavor
    (triangles) QCD as a function of up and down quark masses.
  }
  \label{fig:chit}
\end{figure}

Sea quark mass dependence of $\chi_t$ is plotted in Figure~\ref{fig:chit}.
We observe a good agreement with the expectation from the chiral
effective theory at the leading order,
$\chi_t=\Sigma/(1/m_u+1/m_d+1/m_s)$ \cite{Leutwyler:1992yt}.
($m_s$ is sent to infinity for the two-flavor case.)
It provides another method to extract $\Sigma$.
Our result is $[247(3)(2)\mbox{~MeV}]^3$ (up to the error due to the
scale setting), which is in agreement with the determination from the
spectral function.

\subsection{Vacuum polarization functions}

The vacuum polarization function $\Pi_J^{(0,1)}(Q^2)$ defined through
\begin{equation}
  \int d^4x e^{iqx}\langle 0|J_\mu(x)J_\nu^\dagger(0)|0\rangle
  =
  (g_\mu q^2-q_\mu q_\nu) \Pi_J^{(1)}(Q^2) - q_\mu q_\nu \Pi_J^{(0)}(Q^2)
\end{equation}
for vector ($J=V$) or axial-vector ($J=A$) current is another probe
of the spontaneous symmetry breaking.
For example, the difference between vector and axial-vector channels
is related to $f_\pi$ and $L_{10}$ (one of the NLO low energy
constants in ChPT) as 
\begin{eqnarray}
  \label{eq:Weinberg_SR}
  f_\pi^2 & = & -\lim_{Q^2\to 0} Q^2
  \left[\Pi_V^{(1+0)}(Q^2)-\Pi_A^{(1+0)}(Q^2)\right], \\
  L_{10} & = & -\lim_{Q^2\to 0} \frac{\partial}{\partial Q^2} Q^2
  \left[\Pi_V^{(1+0)}(Q^2)-\Pi_A^{(1+0)}(Q^2)\right], 
\end{eqnarray}
which are called the Weinberg sum rules \cite{Weinberg:1967kj}.
Another interesting quantity is the electromagnetic mass difference of
pion, which is expressed as \cite{Das:1967it}
\begin{equation}
  \label{eq:DGMLY}
  \Delta m_\pi^2 = - \frac{3\alpha_{EM}}{4\pi f_\pi^2}
  \int_0^\infty dQ^2 Q^2 \left[\Pi_V^{(1+0)}(Q^2)-\Pi_A^{(1+0)}(Q^2)\right],
\end{equation}
in the limit of massless pion.
Since the combination $\langle VV-AA\rangle$ vanishes unless the chiral
symmetry is broken, these quantities signal the chiral symmetry
breaking. 

\begin{figure}[tb]
  \centering
  \includegraphics*[width=7cm]{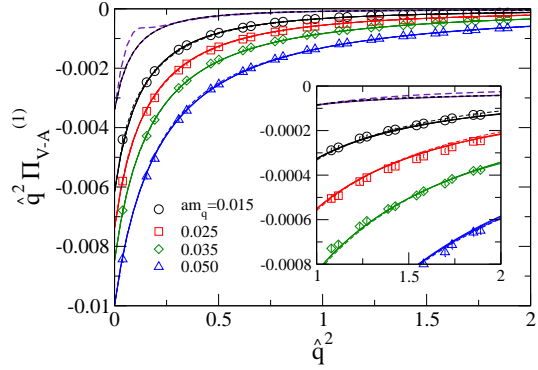}
  \caption{
    Difference of the vacuum polarization functions,
    $\Pi_V^{(1)}(Q^2)-\Pi_A^{(1)}(Q^2)$, multiplied by $Q^2$.
    Data are shown for different quark masses; the chiral limit is also
    plotted by solid (linear extrapolation) and dashed (chiral logs)
    curves. 
  }
  \label{fig:v-a}
\end{figure}

For the lattice calculation of these quantities, exact chiral symmetry
is essential, since we have to extract a tiny difference between the
vector and axial-vector channels.
We extracted the difference 
$\Pi_V^{(1)}(Q^2)-\Pi_A^{(1)}(Q^2)$
successfully, using the overlap fermion on the lattice
\cite{Shintani:2008qe}, as shown in Figure~\ref{fig:v-a}.
The intercept and slope of $Q^2[\Pi_V^{(1)}(Q^2)-\Pi_A^{(1)}(Q^2)]$ at
$Q^2=0$ correspond to $f_\pi^2$ and $L_{10}$, respectively, 
and the integral over $Q^2$ gives $\Delta m_\pi^2$.
The results clearly show that the spontaneous symmetry breaking
induces these physical quantities as expected.
By further improving the numerical data especially in the low $Q^2$
region, we will be able to precisely extract these quantities.

Another use of the vacuum polarization function is an extraction of the strong
coupling constant by matching the lattice data with the OPE expression in the
perturbative regime \cite{Shintani:2008ga}.

\subsection{Pion mass and decay constant}
ChPT is organized as an expansion in terms of small $m_\pi^2$ and
$p^2$, but the region of convergence of this chiral expansion is not
known a priori. 
Using lattice QCD, one can test the expansion and identify the region of
convergence. 
With the exact chiral symmetry, the test is conceptually cleanest, since no
additional terms to describe the violation of chiral symmetry has to be
introduced.
(With other fermion formulations, this is not the case. The unknown correction
terms are often simply ignored.)

For the pion mass $m_\pi$ and decay constant $f_\pi$, the expansion is given as
\begin{eqnarray}
  \label{eq:chiral_exp_mpi}
  \frac{m_\pi^2}{m_q} & = & 2B
  \left[ 1 + \frac{1}{2} x\ln x + c_3 x + O(x^2) \right],
  \\
  \label{eq:chiral_exp_fpi}
  f_\pi & = & f
  \left[ 1 - x\ln x + c_4 x + O(x^2) \right],
\end{eqnarray}
where $m_\pi$ and $f_\pi$ denote the quantities after the corrections while 
$m$ and $f$ are them at the leading order.
The expansions (\ref{eq:chiral_exp_mpi}) and (\ref{eq:chiral_exp_fpi}) may be
written in terms of either
$x\equiv 2m^2/(4\pi f)^2$, 
$\hat{x}\equiv 2m_\pi^2/(4\pi f)^2$, or
$\xi \equiv 2m_\pi^2/(4\pi f_\pi)^2$
(we use a notation of $f_\pi$ = 131~MeV).
They all give an equivalent description at this order,
while the convergence behavior may depend on the expansion parameter.

\begin{figure}[tb]
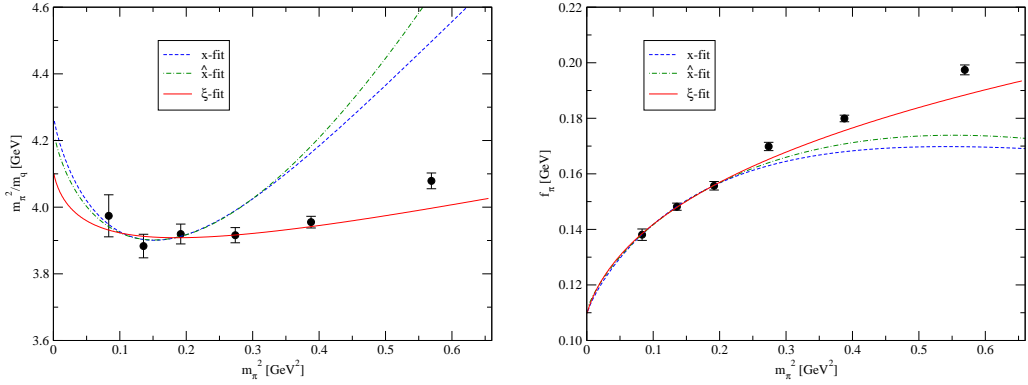

  \centering
  \includegraphics*[width=6.5cm]{figure/mp2r_nlo.eps}
  \hspace*{4mm}
  \includegraphics*[width=6.5cm]{figure/fps_nlo.eps}
  \caption{
    Comparison of chiral expansion in terms of $x$, $\hat{x}$ and
    $\xi$.
    The plots represent $m_\pi^2/m_q$ (left) and $f_\pi$ (right).
    Fits of the three lightest data points with the NLO ChPT formulae
    (\protect\ref{eq:chiral_exp_mpi}) and (\protect\ref{eq:chiral_exp_fpi})
    are shown. 
  } 
  \label{fig:mpifpi}
\end{figure}

Figure~\ref{fig:mpifpi} shows the comparison of different expansion
parameters \cite{Noaki:2008iy} in two-flavor QCD.
The fit curves are obtained by fitting three lightest data points with the
three expansion parameters, which provide equally precise
description of the data in the region of the fit. 
If we look at the heavier quark mass region, however, it is clear that only
the $\xi$-expansion gives a reasonable function and others miss the data
points largely.
This clearly demonstrates that at least for these quantities the convergence
of the chiral expansion is much better with the $\xi$-parameter
than with the other conventional choices.
This is understood as an effect of resummation of the chiral expansion
by the use of the ``renormalized'' quantities $m_\pi^2$ and $f_\pi$.
In fact, only with the $\xi$-expansion we could fit the data including
the kaon mass region with the next-to-next-to-leading (NNLO) formulae
\cite{Noaki:2008iy}.

Whether or not ChPT can be used for kaon is an important question and
potentially has a strong impact on the light hadron phenomenology.
To investigate this question, we are currently extending the analysis
to 2+1-flavor QCD \cite{Noaki:2008gx,JLQCD:2009sk}.

\section{Conclusions}
In this talk, I demonstrate that the scenario of the spontaneous chiral
symmetry breaking in the QCD vacuum is confirmed by lattice QCD
simulations with exact chiral symmetry.
In the calculation of the Dirac operator eigenvalue spectrum, the
expectations from ChPT is precisely tested including the NLO effects. 

With the use of exact chiral symmetry, new possibilities to
extract physics from lattice QCD calculations have been opened.
They include the precise calculation of the topological
susceptibility, the vacuum polarization functions, and a theoretically
clean test of chiral expansion in pion mass and decay constant.
From the project, there are several other interesting calculations of
various physical quantities, such as 
$B_K$ \cite{Aoki:2008ss}, 
meson correlators in the $\epsilon$-regime \cite{Fukaya:2007pn},
nucleon $\sigma$-term and strange quark content \cite{Ohki:2008ff},
and pion form factors \cite{Aoki:2009qn}.

Some of the applications discussed in this talk have been left without
so much progress since the early days of QCD (or even before).
At last, numerical simulation of lattice QCD has caught up 
theoretical conjectures made in 1960s and 70s, but now starting from
the first-principles. 

\vspace*{4mm}
I thank the members of JLQCD/TWQCD for fruitful collaborations. 
The author is supported in part by Grant-in-aid for Scientific
Research (No.~21674002).

\end{document}